%% file: main.tex
\documentclass[10pt, conference, letterpaper]{IEEEtran}

\usepackage{amssymb}
\usepackage{multirow}
\usepackage{amsfonts}
\usepackage{amsmath}
\usepackage{graphicx}
\usepackage{nomencl}
\usepackage{url}
\usepackage{array,booktabs,tabularx}
\usepackage{changes}
\usepackage{array}
\usepackage{rotating}
\usepackage{diagbox}
\usepackage{pifont}
\usepackage{adjustbox}
\newcommand{\cmark}{\ding{51}}%
\newcommand{\xmark}{\ding{55}}
\newcolumntype{C}{>{\centering\arraybackslash}X}

\begin{document}

\title{ThingPot: an interactive Internet-of-Things honeypot} 

\author{

\IEEEauthorblockN{Meng~Wang\IEEEauthorrefmark{1}, Javier~Santillan\IEEEauthorrefmark{2}, and Fernando~Kuipers\IEEEauthorrefmark{1}}

\IEEEauthorblockA{\IEEEauthorrefmark{1}Delft University of Technology\\
2628 CD, Delft, The Netherlands\\
Emails: {mengmengada@gmail.com, F.A.Kuipers@tudelft.nl}}

\IEEEauthorblockA{\IEEEauthorrefmark{2}Brightsight\\
2628 XJ, Delft, The Netherlands\\
Email: jusafing@jusanet.org}

}

\maketitle

\input{abstract}

\input{chapter-1}

\input{chapter-2}
\input{chapter-3}
\input{chapter-4}
\input{chapter-5}
\input{chapter-6}
\bibliographystyle{plain}
\bibliography{report.bib} 


  \end{document}

%% file: abstract.tex
\begin{abstract}
The Mirai Distributed Denial-of-Service (DDoS) attack exploited security vulnerabilities of Internet-of-Things (IoT) devices and thereby clearly signaled that attackers have IoT on their radar. Securing IoT is therefore imperative, but in order to do so it is crucial to understand the strategies of such attackers. 
For that purpose, in this paper, 
a novel IoT honeypot called ThingPot is proposed and deployed. 
Honeypot technology mimics devices that might be exploited by attackers and logs their behavior to detect and analyze the used attack vectors.

ThingPot is the first of its kind, since it focuses not only on the IoT application protocols themselves, but on the whole IoT platform. A Proof-of-Concept is implemented with XMPP and a REST API, to mimic a Philips Hue smart lighting system. ThingPot has been deployed for 1.5 months and through the captured data we have found five types of attacks and attack vectors against smart devices.
The ThingPot source code is made available as open source.
\end{abstract}

%% file: chapter-1.tex
\section{Introduction}
The Internet of Things (IoT) is gaining attention from both industry and society. It is predicted that by 2020 tens of billions of IoT devices will be deployed worldwide, creating a wealth of data \cite{Two}. 
IoT devices often have limited resources, 
posing new stringent requirements for IoT communications. A plethora of protocols have been used and developed to adhere to the IoT requirements. For example, the application-layer protocols XMPP, CoAP, and HTTP are widely used in IoT products. However, these protocols were designed for other purposes, such as real-time communication, asynchronous communication, and messaging, or they are new and possibly not well tested. In either case, it is essential to understand their security vulnerabilities, such that we can attempt to fix them.

Our approach to gaining such an understanding is to develop, to the best of our knowledge, the first interactive IoT honeypot, called ThingPot, which simulates an entire IoT platform, rather than a single application-layer communication protocol (e.g., Telnet, HTTP, etc.). 
A honeypot is a system that captures and identifies malicious activities by simulating a real system or protocol. It is intended to be attacked, but since it is placed in a controlled environment any attacks will be contained. This technology was initially proposed in the late 90's by The Honeynet Project \cite{thehoneynetproject}, and continuously developed since then by the IT security community evolving into more complex deception techniques. In this way, the attacker does not know (at least not initially) that the honeypot is not a real system or device and would try to exploit it based on known vulnerabilities. The attack strategies are recorded by the honeypot, and may include network traffic, payload, malware samples, toolkit used by the attacker, etc. 
Currently, a few IoT honeypots exist:
\begin{itemize}
\item Telnet IoT honeypot \cite{telnetpythonhoneypot}: This is a honeypot that implements a Telnet server to catch IoT malware.
\item HoneyThing \cite{HoneyThing}: This is a honeypot that is designed for TR-069 (CPE WAN Management Protocol).
\item IoTPOT \cite{eleven}: A honeypot to emulate Telnet services of various IoT devices.  
IoTPOT consists of a frontend low-interaction responder cooperating with a backend high-interaction virtual environment called IoTBOX. IoTBOX operates various virtual environments commonly used by embedded systems for different CPU architectures.
\item Dionaea \cite{Dioneagit} \cite{dionaea-orig}: A honeypot framework that, among others, implements an MQTT module. 
\item ZigBee Honeypot \cite{zigbeehoneypot}: A honeypot that simulates a ZigBee gateway.
\item Multi-purpose IoT honeypot \cite{honeypotthesis}: An IoT honeypot that focuses on Telnet, SSH, HTTP, and CWMP. 
\end{itemize}

A comprehensive honeypot that emulates the IoT platform is currently missing. This gap is filled by ThingPot.

The remainder of this paper is structured as follows: Section \ref{chapter2} introduces several popular IoT application protocols and their security mechanisms. In Section \ref{chapter3}, we elaborate on honeypot technology and present the design of ThingPot, our IoT platform honeypot.  
Our ThingPot implementation has been operated in the wild. Section \ref{chapter5} gives our results from the data collected by ThingPot. We conclude in Section \ref{chapter6}.

%% file: chapter-2.tex
\section{IoT application-layer protocols}\label{chapter2}

Already many application-layer protocols have been used for IoT communication, some of the more popular ones are presented in the following:
\begin{itemize}
\item \textbf{MQTT: Message Queue Telemetry Transport}\\
MQTT is a messaging protocol and was released by IBM in 1999. 
It uses the publish/subscribe pattern that runs on top of TCP, so the client does not require updates. 
\item \textbf{XMPP: Extensible Messaging and Presence Protocol}\\
XMPP is a communication protocol that provides basic instant messaging (IM) and presence functionality. It was standardized by the IETF and has been widely used.  
XMPP is extensible, since it allows the specification of XMPP Extension Protocols (XEP) to increase functionality. New XEPs have been released to support IoT. 
\item \textbf{AMQP: Advanced Message Queuing Protocol}\\
AMQP is an open-standard application-layer protocol that arose from the financial industry. It uses a publish/subscribe communication model and supports reliable communication. 
\item \textbf{CoAP: Constrained Application Protocol}\\
CoAP was designed for resource-constrained devices. It is a request/response protocol that runs over UDP. 
It supports QoS and uses a simple Stop-and-Wait retransmission mechanism for confirmed messages \cite{Three}. 
\item \textbf{UPnP: Universal Plug and Play}\\
UPnP is a set of network protocols that are used for the discovery of network devices.  
It is a distributed, open-architecture protocol, based on established standards such as TCP/IP, HTTP, XML, and SOAP. 
\item \textbf{JMS: Java Message Service}\\
JMS is an Application Programming Interface (API) for communication between applications or distributed systems.  

\item \textbf{HTTP REST}\\
REST is an architectural style that was developed in 2000 \cite{three-14}. It has been widely used in Machine-to-Machine (M2M) communications and IoT platforms. It provides a resource-oriented messaging system, where the resources are accessible via URI and GET requests, and inputs are accepted via PUT commands \cite{Two}, \cite{Three}.

\item \textbf{DDS: Data Distribution Service}\\
DDS is a middleware protocol from the Object Management Group (OMG) \cite{FIve-76} that lies between the operating system and applications. It uses a publish-subscribe scheme. 
\end{itemize}

Table \ref{comparision} summarizes the main properties of the aforementioned protocols. 
\begin{table*}[ht]
\centering
\newsavebox{\tablebox}
\begin{lrbox}{\tablebox}
\begin{tabular}{llllllll}
\hline
\    &  MQTT   & XMPP & AMQP& CoAP& UPnP &JMS &HTTP REST  \\
\hline
Transport & TCP & TCP & TCP & UDP & UDP &Not specified & HTTP \\

QoS Options&Yes&No&Yes& No & Yes &Addressed by W3C WG&No    \\

 \multirow{2}{*}{\centering Architecture} & \multirow{2}{*}{\centering Pub\slash Sub} & Pub\slash Sub &\multirow{2}{*}{\centering Pub\slash Sub}&\multirow{2}{*}{\centering Req\slash Res}&\multirow{2}{*}{ }& \multirow{2}{*}{\centering Req\slash Res} &\multirow{2}{*}{\centering Req\slash Res}\\ && or Req\slash Res &&&&\\
 
Real-time & No & Near Real-time & No  & Near Real-time & No & No & No \\

\hline
\end{tabular}
\end{lrbox}
\scalebox{0.75}{\usebox{\tablebox}}
\caption{Properties of popular IoT communication protocols.}
\label{comparision}
\end{table*} 
Table \ref{security} lists their security mechanisms.
\begin{table}[ht]
\centering
\begin{lrbox}{\tablebox}
\begin{tabular}{l|l}
\hline
  MQTT   & Simple User-name\slash password Authentication, TLS\slash SSL for data encryption  \\ 
XMPP & SASL authentication, TLS\slash SSL for data encryption \\

AMQP& SASL authentication, TLS\slash SSL for data encryption   \\
 
CoAP & DTLS\slash IPSEC \\
JMS & Vendor specific but typically based on TLS\slash SSL. Commonly used with JAAS API\\
SOAP & Address by WS-Security\\
REST & TLS\slash SSL plus application server authentication (HTTP server, AAA scheme, etc).\\
\hline
\end{tabular}
\end{lrbox}
\scalebox{0.75}{\usebox{\tablebox}}
\caption{Security mechanisms of popular IoT communications protocols.}
\label{security}
\end{table}
 
\begin{figure}[ht]
    \centering    \includegraphics[width=5cm]{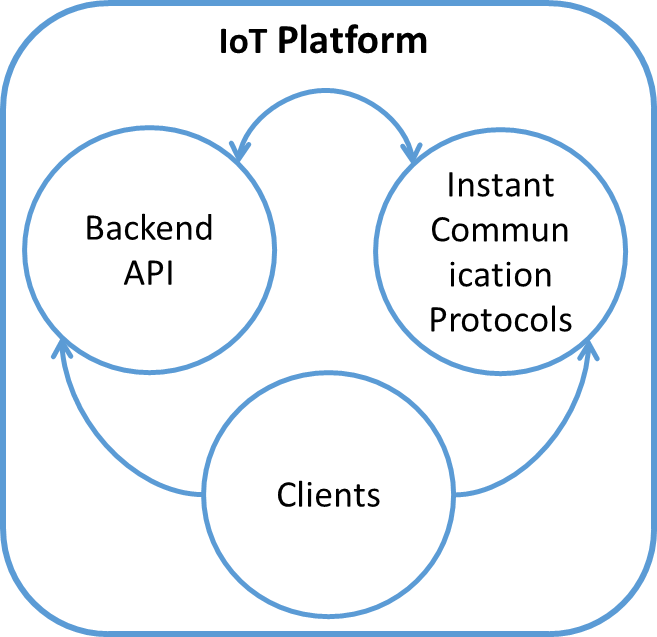}
    \caption{Representative framework of an IoT Platform.}
    \label{currentIoT}
\end{figure}

A representative IoT platform framework is shown in Figure \ref{currentIoT}. It includes an API to work and communicate with the devices, instant communication protocols to communicate between users and API, and the clients (users) that can reach both via the API and instant communication protocols. The individual protocols already might have some security vulnerabilities, which may be augmented in a complex environment were multiple protocols are integrated. 

%% file: chapter-3.tex
\section{ThingPot design and implementation}\label{chapter3}
\begin{figure}[ht]
    \centering    \includegraphics[width=7cm]{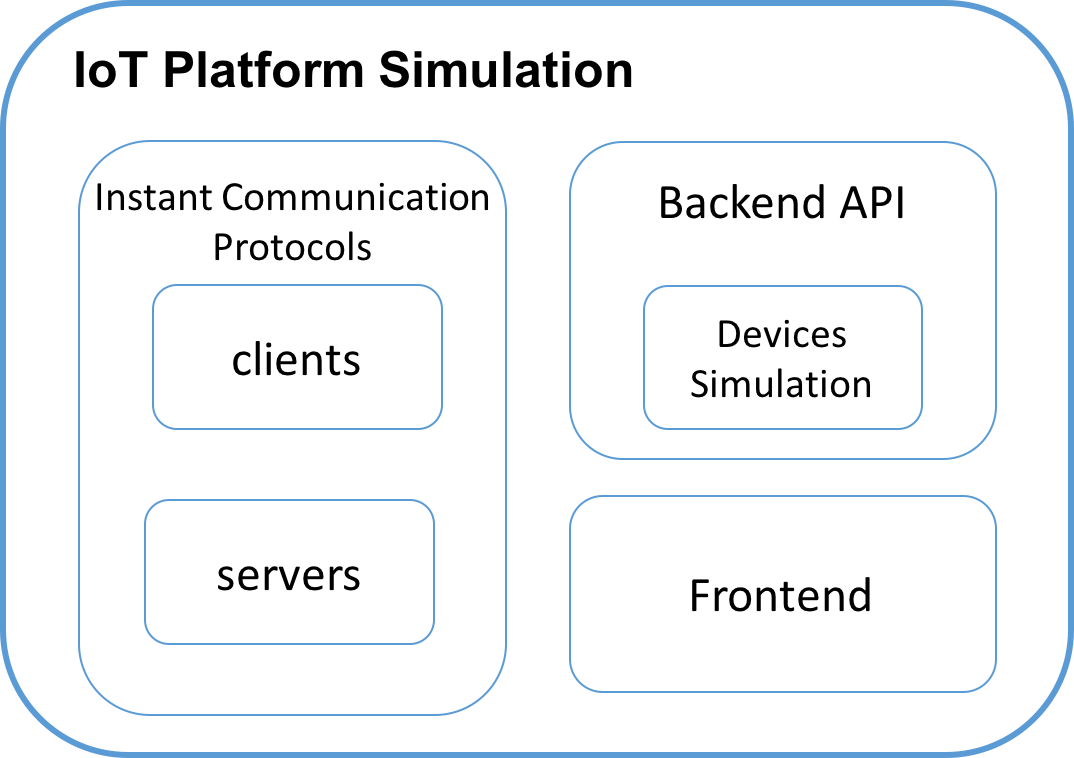}
    \caption{IoT Platform Simulation}
    \label{iotplat}
\end{figure}

In this section, ThingPot, our interactive IoT honeypot, will be introduced. The open-source code of ThingPot is available at \cite{thingpot}. 
In general, based on the level of interaction, one could classify honeypots in three categories: (1) High Interaction Honeypots (HIH): deployment of real systems; (2) Low Interaction Honeypots (LIH): emulation of a system or protocol; and (3) Medium Interaction Honeypots (MIH): combination of both. ThingPot can be considered a MIH or hybrid interaction IoT platform honeypot, whose platform comprises XMPP/MQTT as HIH modules, while LIH device emulation is done through a REST API.  
ThingPot simulates the frontend, backend, IoT devices, and existing XMPP/MQTT services (servers, clients, libraries), see Figure \ref{iotplat}. All of these components compose the IoT platform that hackers could interact with.

%% file: chapter-4.tex
\subsection{ThingPot implementation}
\begin{figure}[ht]
    \centering    \includegraphics[width=8cm]{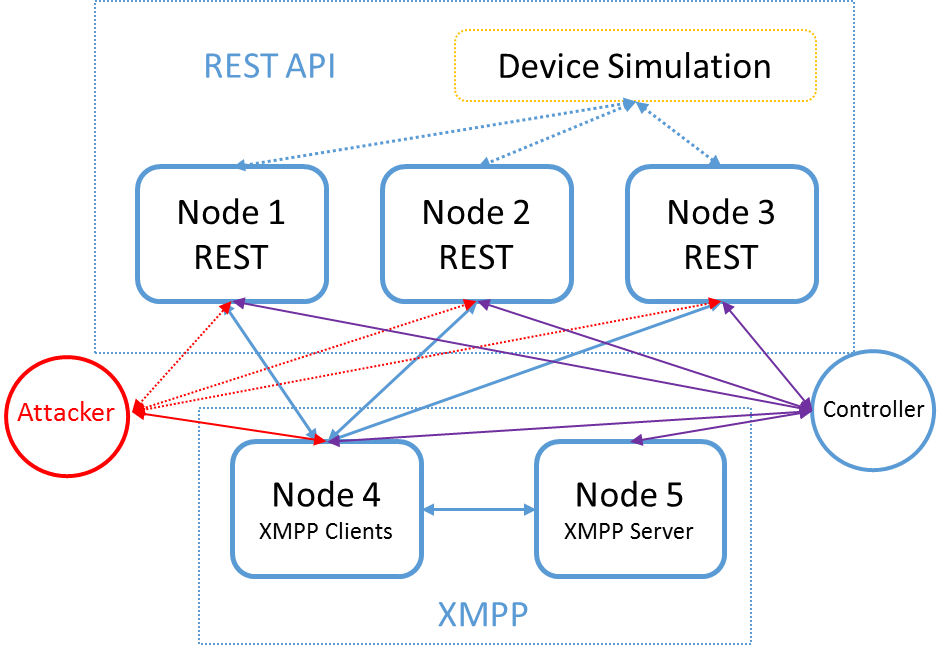}
    \caption{Physical topology of our ThingPot PoC.}
    \label{PicPhysicaltopo}
\end{figure}

REST is used to build a backend API. Given its popularity, we have selected XMPP as the IoT protocol for real-time communication. The frontend is realized through a simple HTTP web service.   
Figure \ref{PicPhysicaltopo} describes the physical topology of our Proof-of-Concept (PoC) implementation of ThingPot. 

Our ThingPot implementation includes the three main components described in Figure \ref{currentIoT}. These components are organized as follows:
\begin{itemize}
\item XMPP nodes:
Two nodes will be used for the XMPP part: one node that runs the XMPP client services and one node that runs the XMPP server services. 
\item REST nodes:
Three nodes with public IP addresses are used to implement the REST part.
\item Controller: 
A computer is used for logging, data storage, updating of the code, etc. 
\end{itemize}

The services are running within an isolated (virtual) environment 
to decrease the chance that the attacker can reach the actual system behind the service emulation of ThingPot. Moreover, on nodes 1 and 2, proxies are running as a backup countermeasure in case the API is temporarily down, as well as to provide a masking mechanism to hide certain information about the backend API. The XMPP server could be either a public server (e.g., xmpp.jp) or a self-hosted server.  

\begin{figure*}[ht]
    \centering    \includegraphics[width=13cm]{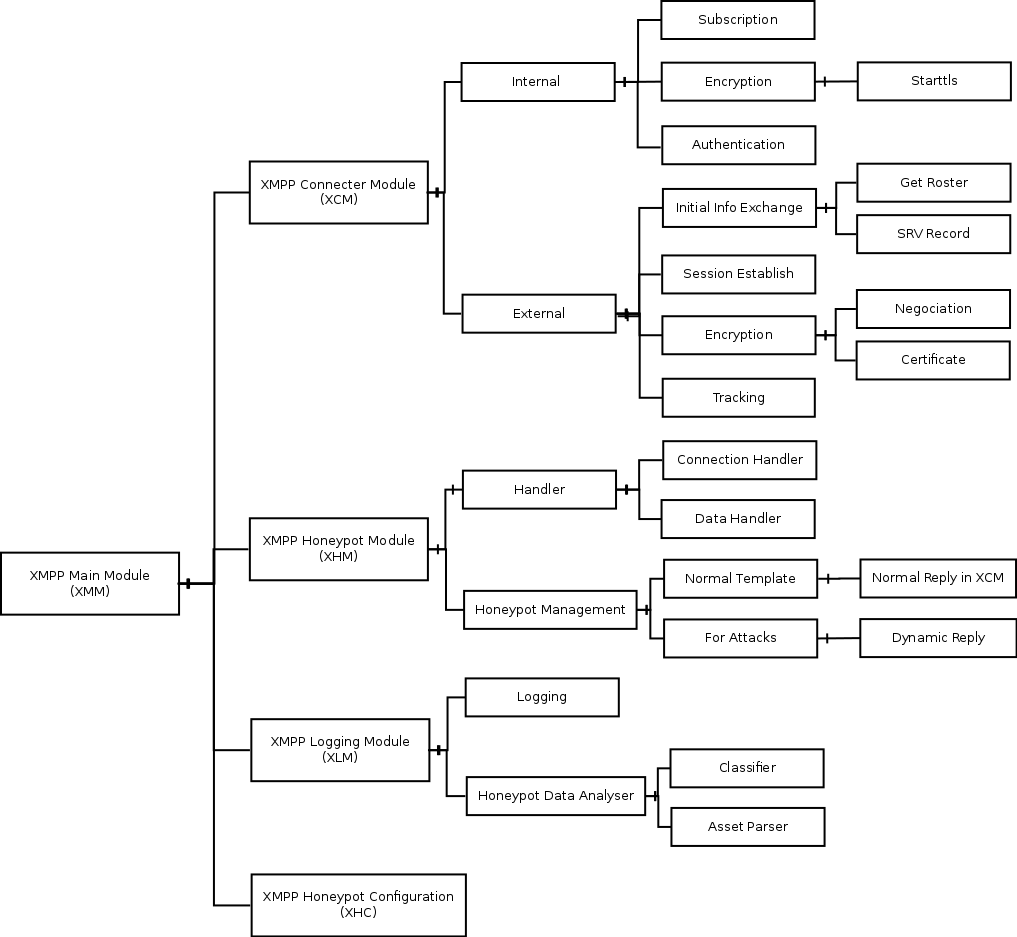}
    \caption{Block diagram of our XMPP client.}
    \label{blockdia}
\end{figure*}

Figure \ref{blockdia} shows a block diagram of our XMPP client component. Table \ref{tabNodes} shows the IP addresses and services on each node. Nodes 1 and 5 are running on a Raspberry Pi (RPi) deployed on different (docker) containers that can be reached via different TCP ports. Nodes 2 and 3 are independent instances running on different servers. 

The left block (in red) of Figure \ref{PicPhysicaltopo} shows a potential attacker. The upper part (in yellow) shows the ``device'' behind the API.   
Through a specific port (see Table \ref{tabNodes}) the attacker can reach and interact with the honeypot. 
No public IP address is needed for the XMPP client, since it only needs to connect to the XMPP server to subscribe emulated devices. 

\begin{table}[ht]
\centering
\begin{tabular}{llll}
\hline
  Node  & Service  & IP address & Port  \\ 
  \hline
1  & REST& 94.210.X.X & TCP/80 \\
2& REST & 83.84.X.X  & TCP/80\\
3 & REST & 84.19.X.X  & TCP/80 \\
4 & XMPP client & internal server & - \\
5 & XMPP server & 94.210.X.X & TCP/5222\&5269 \\
\hline
\end{tabular}
\caption{IP addresses of the nodes in our PoC.}
\label{tabNodes}
\end{table}

\begin{figure}[ht]
    \centering    \includegraphics[width=8cm]{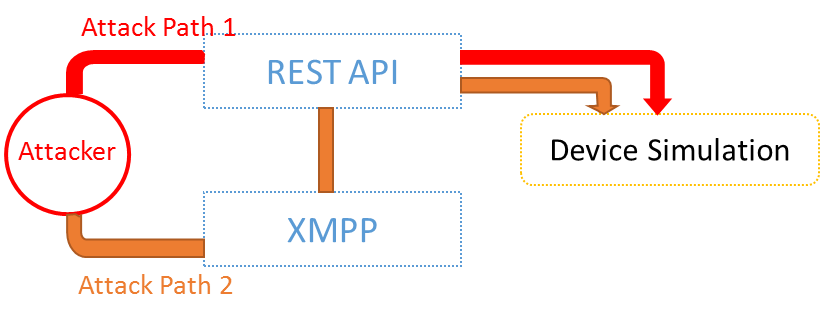}
    \caption{Attack paths.}
    \label{attackpath}
\end{figure}

In this simple, yet general, IoT architecture, we can identify two main paths that the attacker could take to reach the ``device'', as shown in Figure \ref{attackpath}. In the next subsection, we will introduce the practical use case that we had ThingPot simulate, to see which attack paths are used most in that use case.
\subsection{Use case: Philips Hue (smart lighting system)}\label{sectionPhue}
In order to demonstrate the functioning of ThingPot, we chose to mimic an existing IoT product, namely Philips Hue (Phue). 

\begin{figure}[ht]
    \centering    \includegraphics[width=8cm]{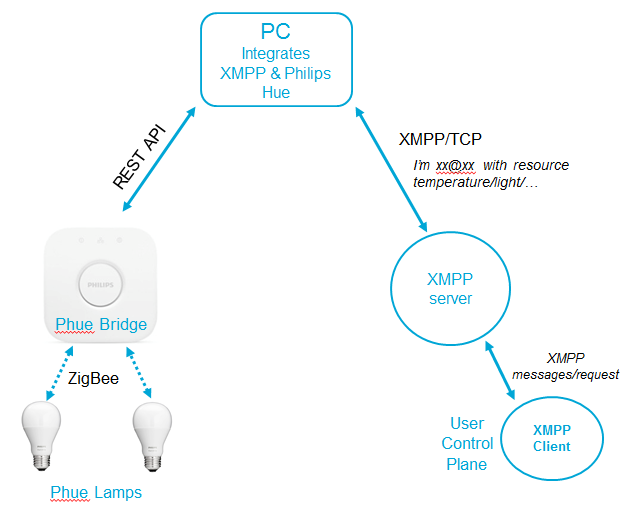}
    \caption{Philips Hue \& XMPP integration.}
    \label{topology}
\end{figure}

Phue consists of wireless LED light bulbs and a wireless bridge. The bulbs can be controlled using iOS and Android apps or through the Meethue website \cite{meethue}, which is a web frontend that communicates with the devices connected to the bridge.  
Third-party implementations (source code) and technical manuals \cite{phue-iot} are available to explain how XMPP can be used with Phue, which is what ThingPot will simulate, 
see Figure \ref{topology}.  
Four main parts are involved:
\begin{itemize}
\item XMPP server: The XMPP server is for the XMPP communications. It transfers all the messages between the XMPP clients. 
\item Phue devices: This part includes the Phue Bridge (a hub) and the smart lamps.  
The Phue Bridge communicates wirelessly with the smart devices through \textit{ZigBee}, whereas the communication with the Phue server and users frontend is performed through a REST API. The Phue bridge has all the information of the connected devices.  
\item PC-integrator of Phue and XMPP: \cite{phue-iot} introduced a way to integrate XMPP with Phue devices. Through this integrator each smart device can have a \textit{JID}, such that the user can send messages to the \textit{JID} to control the lights. This integrator is a script that uses the \textit{sleekxmpp} and \textit{Phue} libraries. 
\item XMPP client: Users can apply for an XMPP account on any public XMPP server and then communicate with the \textit{Phue} devices through XMPP. 
\end{itemize}

\subsubsection{Simulated scenario}
One feature of a honeypot is ``baiting'', which means the honeypot should convince the attacker to believe that it is a real system with vulnerabilities. 

Our simulated scenario is a Philips Hue system that has two smart bulbs connecting to one Phue bridge.  
The Phue bridge API has a public IP address. An XMPP client is used to communicate with the server. A JID has been registered on an XMPP server for controlling and monitoring one smart bulb.  
\subsubsection{REST implementation} 
Our ThingPot PoC contains a RESTful API that simulates the behavior of the Phue bridge API. The \textit{DjangoREST} framework is used to build the API. Figure \ref{figRESTBlock} describes the block diagram of the REST module of the honeypot.  
From the block diagram we can see that the honeypot REST is divided into two parts, the \textit{API} and the \textit{manager}. This structure is based on the Django design, where a \textit{manager} is used for managing the API and the \textit{API} realizes the API. It harbors 6 components:

\begin{figure}[ht]
    \centering    \includegraphics[width=8cm]{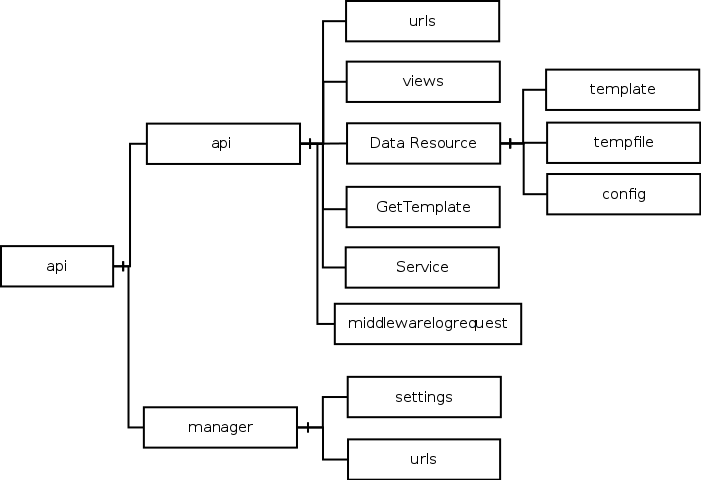}
    \caption{Block diagram of the honeypot REST implementation.}
    \label{figRESTBlock}
\end{figure}
\begin{itemize}
\item \textit{URLs}: This  
maps URL patterns (simple regular expressions) to Python functions (\textit{views}).
\item \textit{views}: This is the Python code that takes requests and returns responses. 
\item \textit{Data Resource}: This includes three JSON files that form a template for response. \textit{template} and \textit{config} follow the format of the Phue bridge resource data structure.  
\textit{tempfile} is an additional feature of the honeypot implementation that the real Phue bridge does not have. It is designed to allow the attacker to obtain more information about the ``device''. 
\item \textit{GetTemplate}: This is a class that can be called by \textit{views}. It handles the response to incoming requests, managing what to respond and calling the correct data from the \textit{Data Resource}.
\item \textit{Service}: \textit{Service} is a class for some specific functions, such as generating a random string, parsing the HTTP header, etc.
\item \textit{middlewarelogrequest}: A customized middleware is built and placed under the \textit{API}.  
It parses all the incoming requests and logs information in JSON format.  
Section \ref{sectionlogsys} explains the details of our log system. 
\end{itemize}

\subsubsection{Logging system} \label{sectionlogsys}

Each part (REST and XMPP) of the honeypot has its own logging system.\\ 
\textbf{REST logs}\\
The REST logging system classifies the logs into two classes: (1) normal Django console log and (2) the incoming requests and corresponding responses.  The Django console logs all the events performed by the REST subsystem. An important fact is that the framework stores on the log every single request and its corresponding response, even if the received request is invalid. This is especially useful for further data analysis, since an invalid request might be part of malicious or anomalous activity from the attacker.  The nature of invalid requests can be defined not only by the format of the HTTP request, but also when the payload does not correspond to an expected format used by the Phue specification.\\ 
\textbf{XMPP logs}\\
The XMPP logging system also has two classes of logs. Since the XMPP honeypot is based on \textit{sleekxmpp} and \textit{phue-lib}, a class -- called XMPP system log -- is formed by the logs generated by these libraries, showing the status and information about them.  
The other class of logs is similar to the requests log of the REST honeypot. It is in JSON format and is called the XMPP traffic log, because it stores all the traffic, including the incoming and outgoing messages through XMPP chat or XML stream to the server and the interaction with the API (HTTP request and response).\\
\textbf{The ``shared\_id'' method}\\
The JID on the XMPP client honeypot is linked to the ``smart bulb'' on the REST honeypot, creating some inter-dependency.  
We have adopted a ``shared\_id'' to link the log entries of the REST JSON log and the XMPP JSON log. When receiving a message from the XMPP part, the log system will generate a ``shared\_id'' that is based on the related JID and the current timestamp.  
Then this generated ``shared\_id'' is written in the header of the HTTP request to the API and transferred to the REST honeypot. When the REST honeypot receives this request, the ``shared\_id'' is saved in the JSON entry of the REST JSON log, which allows us to correlate the logs based on the ``shared\_id''.

%% file: chapter-5.tex
\section{Data analysis and results}\label{chapter5}
In this section, we present the data that were collected by our ThingPot.  
ThingPot was running from June 22th till August 7th. During that time, logs from each node were taken.  
The captured requests can be classified as: targeted, untargeted, and undefined. ``Targeted'' means that the request is explicitly directed at the node, rather than a general (untargeted) scanning.  
``Undefined'' means it is not clear whether the request was targeted or not. In total, 113,741 requests were captured. Table \ref{tab-1stclassification} shows the number of targeted, untargeted, and undefined requests. In the column ``Targeted?'', \cmark means that it is a targeted request, \xmark means untargeted, and ``-'' means undefined. 
\begin{table}[!htb]
\centering
\begin{tabular}{lll}
\hline
  Targeted?  & Count & Percentage \\
  \hline
  \cmark & 47,297  &  41.5\%\\
  -      & 10,444  & 9.2\%\\
  \xmark & 56,000 &  49.2\%\\
  \hline
  TOTAL  & 113,741\\
\hline
\end{tabular}
\caption{Distribution of requests classified by targeted or not.}
\label{tab-1stclassification}
\end{table}
We can see that around half of the requests were targeted. 

Most requests were HTTP REST requests, see Table \ref{tab-2ndclassification}.  
The requests starting with ``/api'' are likely targeted requests, because all the valid URLs that are defined by the API honeypot start with ``/api''. 
\begin{table}[!htb]
\centering
\begin{tabular}{lll}
\hline
  ``/api''?  & Count & Percentage \\
  \hline
  \cmark & 48705  &  42.9\%\\
  \xmark & 64760 &  57.1\%\\
  \hline
  TOTAL  & 113465\\
\hline
\end{tabular}
\caption{Distribution of HTTP REST requests.}
\label{tab-2ndclassification}
\end{table} 
By observing possible correlations between URL, IP addresses, request type, request content (body), user agent, and HTTP status code, we might be able to understand the attacks and recognize an attack pattern. 

The user agent can be considered a unique id for the data caught. Each user agent might behave differently, but requests from the same user agents likely display a similar behavior. Correlating the user agent with other values could be helpful to understand the attacks.
Our logs reveal that many requests came from user agents who resemble a web browser.  
It could be that the requests originated from some plugins of a web browser, or the attacker was using a web browser to hide his real user agent.

Table \ref{tab-useragent-type-ip} shows the correlation between user agent, the request type, and number of IPs. Each line reflects one user agent.  
\begin{table*}[ht]
\centering
\begin{tabular}{p{28pt}p{190pt}p{55pt}p{45pt}}
\hline
  Count  & User Agent & Type of Request & Number of IPs \\ 
  \hline
48460 & Mozilla/5.0 Jorgee & HEAD: 48460 & 360\\ 
31567 & shooter & POST: 31567 & 92 \\ 
9229 & Mozilla/5.0 SF/2.10b & GET: 9171      PUT: 29                FOO: 29 & 7 \\ 
2984 & botlight &POST: 2984 &20 \\ 
2378 & 000modscan  & POST: 2378     & 12 \\ 

1867 & httpget & GET: 1867 & 7\\ 
622 & Mozilla/5.0 (Windows NT 5.1; rv:32.0) Gecko/20100101 Firefox/31.0 & GET: 622&14\\ 
607& 0000modscan & GET: 607& 12 \\
275 & Mozilla/5.0 (Macintosh; Intel Mac OS X 10.11; rv:47.0) Gecko/20100101 Firefox/47.0 & GET: 275& 2\\ 
248 & ioscan & GET: 248 & 2\\
96 & Python-urllib/2.7 & GET: 96& 3\\
\hline
\end{tabular}
\caption{Top 20 user agents from the proxy logs.}
\label{tab-useragent-type-ip}
\end{table*} 
By analyzing the correlation between user agent and request type and status, as well as checking details (URLs, body) of part of the log, a summary based on each user agent can be given:
\begin{itemize}
    \item Mozilla/5.0 Jorgee:\\
    ``Jorgee'' is malware that tries to make use of the SQL web admin flaws \cite{jorgee}. It appeared 48460 times in total, with 360 IPs involved, from June 28th to August 2nd and occurs daily. Each IP sent around 200 to 400 requests. All the request types were HEAD (which is similar to GET). The URLs had keywords ``db'', ``admin'', ``pma'', ``php'', ``sql'', ``web'', ``database'', and ``my''. The URL was the permutation and combination of these words. This user agent only appeared in the log of node 3. 
    \item shooter:\\
    This user agent might have been created manually. It generated 31567 requests on the honeypot, with 92 IPs involved. TOR technology was used to hide the source IP. The requests dated from July 19th to July 20th and arrived every 1 to 4 seconds. All the requests' URLs were ``/api/'' with the POST method and a specific body content. Figure \ref{shooter} shows a sample log entry.  
From this entry, we can see that the body content has a well organized format that follows the format of the Philips Hue data structure. A replay of this request was done on a real Philips Hue White. The reply from the real Phue bridge was an error message saying that the parameter is not available.
\begin{figure*}[ht]
    \centering    \includegraphics[width=16cm]{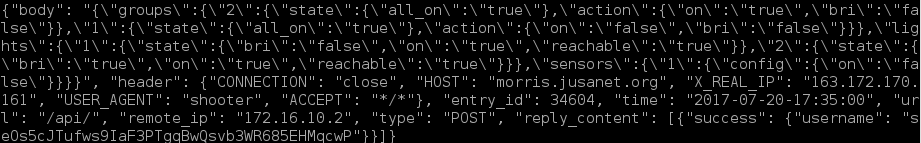}
    \caption{One of the log entries from user agent shooter.}
    \label{shooter}
\end{figure*}
    \item Mozilla/5.0 SF/2.10b:\\
    This is the default user agent for Skipfish \cite{skipfishuseragent}. 7 IPs were involved with 9229 requests in total. The requests were GET, PUT, or FOO. Skipfish is a scanning tool that was designed for security checks, but it could also provide useful information to attackers. 
    \item botlight:\\
    2984 requests were received from this user agent. The burst of requests started at July 19th 14:27 and lasted till 15:50 the same day. All requests were POST requests with the same URL ``/api/list/''. 21 IPs were involved and TOR was used.  
Figure \ref{botlight} shows one log entry from user agent ``botlight''. The body content of the requests is following the \textit{multipart/form-data}, which is used for file upload, but there are also a lot of \%s and 0s, which might be characteristic of fuzzing (a technique that provides invalid or random data as input to find bugs).
\begin{figure*}[ht]
    \centering    \includegraphics[width=16cm]{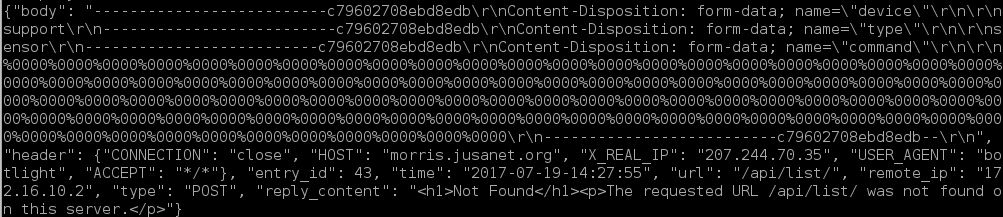}
    \caption{A log entry from user agent botlight.}
    \label{botlight}
\end{figure*}
    \item 000modscan:\\
    000modscan only used POST requests with 12 IPs involved. The requests happened on July 5th from 10:22 till 11:59. The URL contained the patterns \textit{/api/philips/hue/\{32\_chars\}}, \textit{/api/philips2/hue-link/\{32\_chars\}}, \textit{/api/belkin/wemo/\{32\_chars\}}, and \textit{/api/philips1/hue/\{32\_chars\}}. \textit{\{32\_chars\}} means a random 32 long string that contains only digits and lower-case letters. These POST requests had content that was similar to the body of ``botlight''. 
    \item httpget: \\
    The requests from user agent httpget were all HTTP GET requests, with 7 IPs involved. The requests happened in two periods: one on July 5th from 12:02 till 12:18 and the second on July 6th from 11:34 till 11:46. The URLs followed three patterns:  \textit{/api/phi/light/\{32\_chars\}/tokens}, \textit{/api/\{32\_chars\}/tokens}, and \textit{/api/\{32\_chars\}}. Since both days contained the URL pattern \textit{/api/{32\_chars}/tokens}, we assume that the requests originated from one source. 
    \item Mozilla/5.0(WindowsNT5.1;rv:32.0)Gecko/20100101 Firefox/31.0:\\ 
    All the requests were HTTP GET requests, with 14 IPs involved, although most of the requests were from IP 91.196.50.33. The requests dated from June 28th to July 31st, 2 to 5 times per day. The URL of most requests was \textit{/http:/testp3.pospr.waw.pl/testproxy.php}. This is a URL to look for open proxies.
    \item 0000modscan:\\ 
    All the requests from user agent ``0000modscan'' were GET requests, with 12 IPs involved. The requests happened on July 5th from 09:37 to 09:42. The URLs of the requests were \textit{/api/tplink/light/\{32\_chars\}}, \textit{/api/philips/hue/\{32\_chars\}}, \textit{/api/phi/light/\{32\_chars\}}, \textit{/api/philips2/hue-link/\{32\_chars\}}, \textit{/api/belkin/wemo/\{32\_chars\}}, and \textit{/api/philips1/hue/\{32\_chars\}}. This was an HTTP GET scan targeting smart light devices.
    \item Mozilla/5.0   (Macintosh;Intel   Mac   OS   X10.11; rv:47.0) Gecko/20100101 Firefox/47.0:\\
    The requests were only from two IPs: 183.129.160.229 (193 times) and 60.191.38.77 (82 times). All were GET with the URL \textit{/}. The user agent is from a Mac operating system with a Firefox web browser. By checking the time of the requests, they were likely sent by someone who found the proxy and requested the data resource manually.
    \item ioscan:\\
    All the requests from user agent ``ioscan'' were HTTP GET requests, with 2 IPs involved. The requests happened on July 4th from 13:58 till 14:00. All its URLs followed the pattern \textit{/api/hue/\{0-216\}}, where \{0-216\} is a number that ranges from 0 to 216. This can be considered a targeted scan, since ``hue'' is a keyword of the honeypot.
    \item Python-urllib/2.7:\\
    Python-urllib/2.7 is a Python library for fetching data across the World Wide Web \cite{pythonurl}. The requests from user agent ``Python-urllib/2.7'' were all HTTP GET requests, with 3 IPs involved. 94 of the requests happened on July 1st from 01:10:09 till 01:10:19, i.e. 94 requests in 10 seconds, and all came from IP 185.77.172.42. From this we can see that there must have been a script running. 
The URL of these requests seems quite random, but in general was looking for vulnerabilities of the SQL web admin.   
\end{itemize}

\subsection{Findings interpretation}\label{findings}
Table \ref{tab-useragent-type-ip} summarizes our logs and in the following we discuss the main types of attacks found.\\ 
\textbf{Targeted attack that is trying to take control}\\
The attack was an HTTP POST request with a specific HTTP body. The HTTP body was JSON-format data similar to the format of a real reply of the Philips Hue bridge. One example is shown in Figure \ref{shooter}. The instruction of the Philips Hue API says that to control the Hue device, a POST request with a JSON body and a valid URL should be sent to the Philips Hue bridge. This particular attack was simulating such a request to change the value of the Philips Hue bridge. Hence, the attacker already assumed that the implementation was Philips Hue and knew how to control it.\\
\textbf{Attack with the body following the \textit{multipart/form-data} format} \\
This was an attack that used HTTP POST with a specific body content. The body contents from for example ``botlight'' and ``000modscan'' had very similar HTTP POST requests:\\
\texttt{--------------------------\{16\_chars\}\textbackslash r\textbackslash n Content-Disposition: form-data; name=\textbackslash"on
\textbackslash"\textbackslash r\textbackslash n\textbackslash r\textbackslash ntrue\textbackslash r\textbackslash n-------------------------
-\{16\_chars\}\textbackslash r\textbackslash nContent-Disposition: form-
data; name=\textbackslash"productid\textbackslash"\textbackslash r\textbackslash n\textbackslash r\textbackslash n\{random\_pay\\load\}\textbackslash r\textbackslash n--------------------------\{16\_char
s\}--\textbackslash r\textbackslash n}
\\where \{16\_chars\} is a random 16-long string containing only lower-case letters and digits. \{random\_payload\} is something random, which ``botlight'' filled with ``\%0000'' and ``000modscan'' filled with nothing. An other user agent ``mass'' was also sending POST requests with this pattern, and where \{random\_payload\} was filled with an extremely long repeated (9944 times in one payload) ``\%A/telnet'' or ``\%A/xmpp'', or ``\%A/upnp'', apparently a fuzzing attempt.
\textbf{Attack with URL}\\
This kind of attack was through HTTP GET with URLs following a specific pattern:
\begin{enumerate}
    \item /api/philips/hue/\{32\_chars\}
    \item /api/phi/light/\{32\_chars\}
    \item /api/philips1/hue/\{32\_chars\}
    \item /api/philips2/hue-link/\{32\_chars\}
    \item /api/belkin/wemo/\{32\_chars\}
    \item /api/tplink/light/\{32\_chars\}
    \item /api/hue/\{0-750\}
    \item /api/phi/light/\{32\_chars\}/tokens
    \item /api/\{32\_chars\}/tokens
    \item /api/\{32\_chars\}
\end{enumerate}
Where \{32\_chars\} means a 32-long string that contains lower-case letters and digits randomly. And \{0-750\} means a number ranging from 0 to 750. All of these requests were targeted scanning attacks.\\
\textbf{General scanning tools or libraries}\\
A number of scanning tools and libraries appeared in the honeypot data: 
\begin{itemize}
    \item skipfish \cite{skipfishuseragent}
    \item Nikto \cite{nikto}
    \item Jorgee: there is not much information about this, but it appears to be a web scanner. 
    \item masscan \cite{masscan}: This was also found in the user agent list. The source code address is also included in the header.
    \item Python library urllib \cite{pythonurl}
    \item http://testp3.pospr.waw.pl/testproxy.php: It can reveal the IP of the proxy in your network.
    \item Proxyradar: On \textit{https://proxyradar.com/} you can find open proxies.
\end{itemize}
\textbf{Other unrelated attacks}\\
Attacks that were executing commands to find something, but which were not related to the IoT platform were also found. 

%% file: chapter-6.tex
\section{Conclusion}\label{chapter6}
In this paper, 
we have proposed a design of an IoT platform honeypot (ThingPot). Moreover, we have provided a Proof-of-Concept (PoC) implementation of ThingPot, which is made available as open source code, focusing on the IoT platform use case of Philips Hue (smart lights system). Two important components of that ThingPot PoC use case are an XMPP client and a REST API. Our ThingPot PoC has been deployed and captured real data for 1.5 months. 

Our analysis of the captured data shows that there were only few attacker activities (even generic scans or requests) on the XMPP part. In fact, there were no direct requests for the devices through the XMPP path. This supports the statement that XMPP increases the complexity for attackers to reach the IoT platform and devices. Other reasons could be that, currently, attackers are not yet showing interest in exploiting the flaws of XMPP on an IoT platform. Or attacks on XMPP target mainly the XMPP server, which was out of the scope of our ThingPot PoC. 
This does not mean that potential security problems in using XMPP for IoT should not to be taken into consideration. More work on the protection of the XMPP accounts (prevention from being exposed, additional authentication, ...) should be done to reach a more secure XMPP-IoT Platform.  

Our analysis of the data from the REST logs indicates that the IoT platforms and devices have been noticed by attackers. 
In particular, five main kinds of attacks were found. In general, attackers are looking for devices like Philips Hue, Belkin Wemo, TPlink, etc. In particular, they are interested in getting information about the smart devices and to take over control of them. Often, the attackers are using the TOR network to mask their real source. The methodology that the attackers seemed to prefer is first a general scanning to look for openings, followed by a more targeted and specific attack via brute force or fuzzing. An attack specifically targeting the Philips Hue was also found. 